\documentclass[peerreview,draftcls,onecolumn,12pt,a4paper]{IEEEtran}
\usepackage{algorithmic}
\usepackage{algorithm}
\usepackage[dvips]{graphicx}
\usepackage{times}
\usepackage{caption2}
\usepackage{verbatim}
\input epsf
\usepackage[cmex10]{amsmath}
\usepackage{amssymb}
\usepackage{cite}      
\include{epsf}
\usepackage{url}
\usepackage{multirow}
\usepackage{color}
\usepackage{setspace}
\usepackage[tight,footnotesize]{subfigure}
\IEEEoverridecommandlockouts 

\newcommand{\ud}{\,\mathrm{d}}

\newcommand{\ls}[1]  
    {\dimen0=\fontdimen6\the=#1\dimen0
     \advance\lineskip.5\fontdimen5\the\lineskip-\dimen0
     \lineskiplimit=.9\lineskip
     \baselineskip=\lineskip
     \advance\baselineskip\dimen0
     \normallineskip\lineskip
     \normallineskiplimit\lineskiplimit
     \normalbaselineskip\baselineskip
     \ignorespaces
    }

\renewcommand{\baselinestretch}{1.7}
\begin{document}

\title{STiCMAC: A MAC Protocol for Robust Space-Time Coding in Cooperative Wireless LANs}

\author{
\authorblockN{\small  Pei Liu$^{\star}$, Chun Nie$^{\star}$, Thanasis Korakis$^{\star}$, Elza Erkip$^{\star}$, Shivendra Panwar$^{\star}$,
Francesco
Verde $^{\dagger}$, Anna Scaglione$^{\diamond}$}\\
\authorblockA
{\small $^{\star}$ Polytechnic Institute of New York University,
Brooklyn, NY, USA\\} {\small $^{\dagger}$ University of Naples,
Federico II, Italy \\} {\small $^{\diamond}$ University of
California, Davis, USA}
\thanks{This paper is a revised version of a paper with the same name submitted to IEEE Transaction on Wireless Communications. STiCMAC protocol with RTS/CTS turned off is presented in the appendix of this draft.
}
}

\maketitle

\begin{abstract}

Relay-assisted cooperative wireless communication has been shown to have significant performance gains over the legacy direct transmission scheme. Compared with single relay based cooperation schemes, utilizing multiple relays further improves the reliability and rate of transmissions. Distributed space-time coding (DSTC), as one of the schemes to utilize multiple relays, requires tight coordination between relays and does not perform well in a distributed environment with mobility. In this paper, a cooperative medium access control (MAC) layer protocol, called \emph{STiCMAC}, is designed to allow multiple relays to transmit at the same time in an IEEE 802.11 network. The transmission is based on a novel DSTC scheme called \emph{randomized distributed space-time coding} (\emph{R-DSTC}), which requires minimum coordination. Unlike conventional cooperation schemes that pick nodes with good links, \emph{STiCMAC} picks a \emph{transmission mode} that could most improve the end-to-end data rate. Any station that correctly receives from the source can act as a relay and participate in forwarding. The MAC protocol is implemented in a fully decentralized manner and is able to opportunistically recruit relays on the fly, thus making it \emph{robust} to channel variations and user mobility. Simulation results show that the network capacity and delay performance are greatly improved, especially in a mobile environment.
\end{abstract}


\newpage

\vspace{-0.5in}
\section{Introduction}
\label{Introduction}

Cooperative wireless communication~\cite{Sendonaris03a,Sendonaris03b,La03,La04} techniques exploit the broadcast nature of the wireless channel by allowing stations that overhear other transmissions to relay information to the intended destination, thereby yielding higher reliability and throughput than direct transmission. While initial cooperative communication schemes \cite{Sendonaris03a,Sendonaris03b} employ a single relay, subsequent work \cite{La04} allows multiple relays to forward signals at the same time, each mimicking an antenna of a multiple antenna transmitter by using a \emph{distributed space-time code} (DSTC). For a DSTC based transmission scheme, relays must be carefully coordinated. Before each packet transmission, a central node/controller selects and indexes all the relays it wants to recruit. This decision must be known by each of the selected relays, so that they know \emph{who} participates in cooperation and \emph{which} signal stream of the DSTC each of them transmits. In a distributed environment with mobility, this leads to extra signaling overhead. Furthermore, the controller needs global channel knowledge in order to optimize system performance. Another drawback of this scheme is that nodes other than those being chosen are prohibited from relaying, while at the same time, the chosen relays might fail to participate in forwarding the signal due to fading or noise. Those inherent drawbacks lead to inefficiencies in implementing a DSTC-based protocol.

The above drawbacks can be addressed by employing \emph{randomized distributed space-time coding} (R-DSTC)~\cite{sirkeci_scaglione_mergen_2007_SP}, which eliminates the requirement of space-time code (STC) codeword assignment and reduces the coordination between the source and the  relays. R-DSTC provides a robust cooperative relaying scheme in contrast to a DSTC based system, and has the potential of simplifying the protocol design, thus leading to a reduction in signaling cost.

In a cooperative environment, physical (PHY) layer cooperation needs to be integrated with a medium access control (MAC) layer in order to recruit relays as well as coordinate transmissions and receptions. \emph{CoopMAC}~\cite{coopmacliu}, as one of the first MAC layer designs to support a cooperative PHY layer in a wireless LAN (WLAN),  enables cooperation under the IEEE 802.11 framework. Since the low data-rate stations at the edge consume the majority of the channel time, the aggregate throughput is severely degraded \cite{Heusse03}. \emph{CoopMAC} alleviates this problem by allowing transmissions to take place in a two hop manner. As the transmissions over both hops are accomplished at a high rate, a considerable improvement is achieved in the aggregated throughput. The performance of CoopMAC, albeit superior to direct communication, is still limited as it only selects a single relay, which is a disadvantage when it is employed in a fading environment.  While utilizing multiple relays at the PHY layer greatly improves the reliability of transmissions, it remains unclear how such techniques can be employed to deliver significant network capacity gains for a loosely synchronized network such as the IEEE 802.11.

In this paper, we design a robust MAC layer protocol called \emph{STiCMAC} (Space-Time coding for Cooperative MAC), which is compliant with the IEEE 802.11 standard and enables R-DSTC based cooperation. \emph{STiCMAC} allows one to harvest cooperative diversity from multiple nodes in a decentralized manner. In this scheme, if a data packet from the source needs to be relayed, all potential relays listen to the transmission from the source and try to decode. Assuming error detection mechanisms  such as  cyclic redundancy check (CRC) are employed at the relays, only relays that successfully decode the packet forward to the destination in \emph{unison} using R-DSTC. In order to do so, the handshaking procedure defined by the IEEE 802.11 standard is extended to allow relays be recruited in an opportunistic manner while ensuring that the transmissions from multiple relays are collision free.

The main contribution of this paper is that it fundamentally changes the way cooperation is established. Instead of picking nodes with fast links or finding a fast path in the network, our scheme picks a \emph{ transmission mode} (modulation, channel coding and STC) that could most improve the end-to-end rate on the average. Relays decide to participate or not to participate independently based on whether they receive the packet or not. In fact, neither the source or destination station need to know who the relays are or where they are located. STiCMAC is an optimized PHY/MAC cross-layer scheme that can be implemented in a fully \emph{decentralized} manner and is able to opportunistically recruit relays \emph{on the fly} at minimum signaling cost for an infrastructure-based IEEE 802.11 network.

We evaluate the system performance of \emph{STiCMAC}, and employ
cross-layer optimization to find suitable transmission parameters,
i.e., per-hop rates and STC dimension, that maximize the
end-to-end rate.  Optimization of transmission parameters is
performed assuming either a complete knowledge of average channel
statistics for both hops, or simply considering the number of
stations in each WLAN cell. We call these two approaches
\emph{STiCMAC with channel statistics} (\emph{STiCMAC-CS}) and
\emph{STiCMAC with user count} (\emph{STiCMAC-UC}) respectively.
We investigate the aggregated network throughput and the average
delay for all stations in both a static and mobile environment.
Our results suggest that both STiCMAC-CS and STiCMAC-UC have
similar performance, especially for a large number of users. This
outcome strongly supports our argument that the proposed scheme
does not need a priori knowledge of channel conditions, as opposed
to DSTC. Additionally, simulation results show that
both types of \emph{STiCMAC} significantly outperform DSTC in
terms of throughput and delay, due to lower signaling cost, and
also outperform \emph{CoopMAC} and direct transmission due to
increased diversity. While the performance of DSTC and
\emph{CoopMAC} significantly decreases under mobility,
\emph{STiCMAC} is more robust, and in particular \emph{STiCMAC-UC}
shows minimal performance degradation in a mobile environment.
Finally, we conduct a study
of the interference propagated to neighboring wireless LANs for
all transmission schemes. Simulation results show that, for the
same traffic load, the average interference generated by
\emph{STiCMAC} is similar to DSTC, and much less than
\emph{CoopMAC} and direct transmission.

A related work~\cite{Jakllari06} proposes a MAC layer protocol that deploys DSTC in an \emph{ad hoc} network to assist network layer routing. This allows cooperative transmissions from multiple relays, however, its performance and practicality for a mobile network are still limited due to the limitations of DSTC outlined above. Opportunistic routing~\cite{opprouting} is a routing protocol for \emph{ad hoc} networks that allows one node closest to the destination to forward in case multiple nodes receive from the previous hop. Compared with STiCMAC, which  operates in MAC/PHY cross-layer, opportunistic routing operates in the network layer. The other difference is that STiCMAC allows an end-to-end multihop transmission within a single channel access and queuing is not necessary at the relays. STiCMAC also allows signals from multiple relays to be combined coherently in the PHY layer. Another use of the term ``opportunistic'' appears in the cooperative communications literature in~\cite{pathselect}, however, the notion there is to select relays based on instantaneous channel state. Generic MAC protocols for R-DSTC are designed in ~\cite{peiGlobecom08, RcoopMAC}, where transmission parameters are optimized given the bit error rate, assuming no channel coding is employed. Without forward error correction coding, those schemes can cause error propagation by allow relays to forward even if erroneous packets are received. These papers mostly focus on the PHY layer characteristics that enable the use of randomized codes in realistic wireless networks, and do not explicitly investigate MAC layer details. STiCMAC is the first protocol derived from the IEEE 802.11 where practical MAC layer aspects of randomized cooperation are addressed.

We note that synchronization is an important issue for all transmission schemes that allow multiple stations to transmit at the same time on the same frequency. As demonstrated in~\cite{MurphayDSTC}, symbol level synchronization in DSTC based transmission is feasible on a software defined radio platform with commercially available IEEE 802.11 components and a customized FPGA. Thus we believe necessary synchronization for R-DSTC, which is more robust to synchronization errors than DSTC~\cite{Sharp2008}, can also be implemented.

The rest of this paper is outlined as follows. Section
\ref{R-DSTC} introduces the PHY layer background for R-DSTC. In
Section \ref{WLAN-MAC-R-DSTC}, we present the \emph{STiCMAC}
protocol in detail. Section \ref{sec:rateadapt} develops two
opportunistic rate adaptation schemes for \emph{STiCMAC} to
optimize the transmission parameters.
Section~\ref{PerformanceEvaluation} presents the simulation
results and the performance evaluation. Finally, in Section
\ref{Conclusion}, we present conclusions.

\vspace{-0.2in}
\section{R-DSTC Physical Layer Description}
\label{R-DSTC}

An STC is designed to operate over several antennas at the same transmitter station. In contrast, DSTC employs an STC over multiple relays in a distributed manner. When these relays cooperatively forward a signal, each relay corresponds to a specific antenna element of the underlying STC, and transmits a predefined STC encoded stream. The advantage of DSTC lies in its capability to form a virtual MIMO system by using these relays and producing diversity gain, even if each station is only equipped with one antenna. The performance of DSTC and the diversity gain obtained has been studied in \cite{La03}. 

R-DSTC is introduced and examined in
\cite{sirkeci_scaglione_mergen_2007_SP} as a novel form of DSTC.
Like conventional DSTC, R-DSTC is deployed in a cooperative
scenario with multiple relays along with a source and destination
pair and operates over two hops. Although R-DSTC can be employed
using relays with multiple antennas, we assume that each station
is only equipped with a single antenna. The scenario with multiple
antennas per station can be easily extended from the
single-antenna case.

\begin{figure}
    \centering{\epsfxsize=.8\textwidth \epsfbox{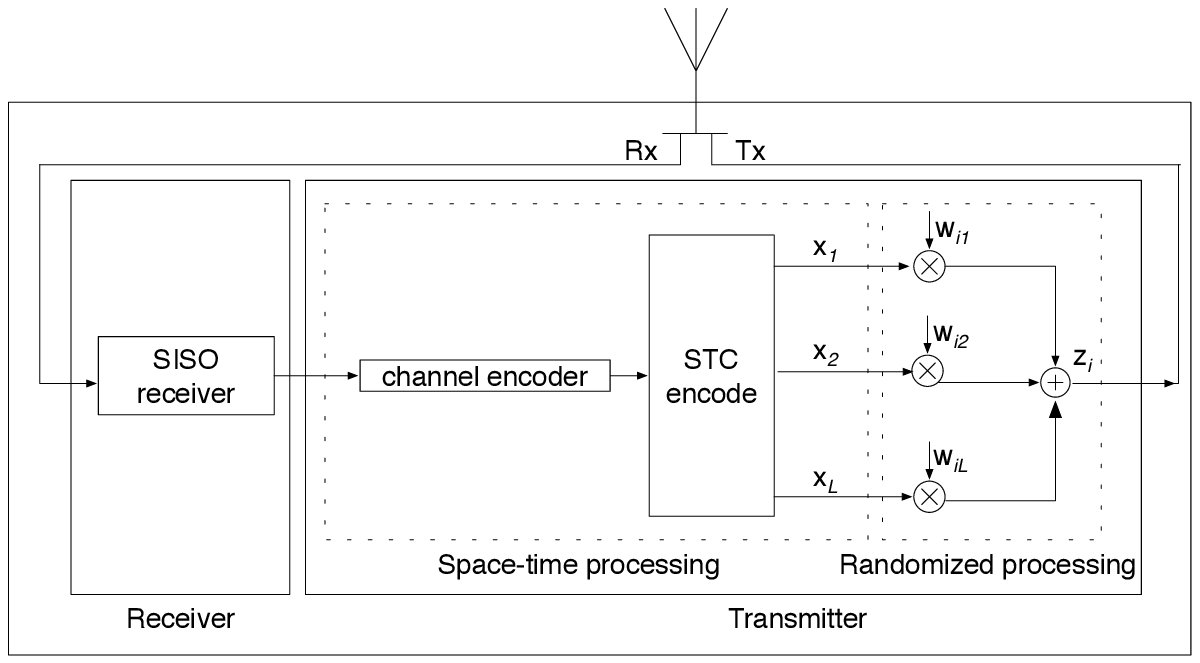}}
    \caption{R-DSTC signal processing in a relay.}
    \label{fig:rdstc_block} \vspace{4in}
\end{figure}

Fig. \ref{fig:rdstc_block} shows a single-antenna relay that
employs a regular single-input and single output (SISO) decoder to
decode the information sent by the source station in the first
hop. Provided the information is decoded correctly, as determined
by checking the CRC field, the relay is responsible for
re-encoding the information bits and passing them to an STC
encoder. Suppose the underlying space-time codeword has a
dimension $L \times K$, where \emph{L} is the number of antennas
and \emph{K} is the block length transmitted by each antenna. The
STC encoder generates an output of $L$ parallel streams, each
stream corresponding to an antenna. Unlike a regular DSTC where
the $j$th relay simply transmits the stream $j$, in a R-DSTC
system the $j$th relay transmits \emph{a linear weighted
combination of all $L$ streams}. The weights of the $L$ streams at
the $j$th relay are denoted by a vector $\mathbf{w}_j=[w_{j1} \;
w_{j2} \; \ldots \; w_{jL}]$. Each element in $\mathbf{w}_j$ is an
independently generated random variable with zero mean and
variance $1/L$. As described in
\cite{sirkeci_scaglione_mergen_2007_SP}, a complex Gaussian
distribution is adopted for the distribution of the weights since
it has desirable properties in terms of PHY layer error rates.
Assuming $n$ relays simultaneously transmit in the second hop, the
vector $\mathbf{w}_j$, where $j=1,2,\dots n$, represents the
random weights at relay $j$ and $\mathbf{R}=[\mathbf{w_1, w_2,
\dots , w_n}]^T$ is the weight matrix for all these $n$ relays.
The destination station is assumed to have one antenna and is able
to decode the received signal with a conventional STC decoder. 

The physical layer fundamentals of R-DSTC is described in detail
in \cite{sirkeci_scaglione_mergen_2007_SP}, where it is shown that
R-DSTC comes very close to the performance of DSTC in terms of PHY
layer properties and can provide the full diversity order of $L$
with at least $L$ relays. A major advantage of the R-DSTC
technique over DSTC lies in the fact that the source station does
not need to specifically select its relays as well as to assign
antenna indices to each relay. In R-DSTC, the source and
destination remain unaware of which stations act as relays and
which random weight vector has been used in relays. These features
enable R-DSTC to be a fully decentralized scheme in a cooperative
environment.

\vspace{-0.2in}
\section{ST\MakeLowercase{i}C MAC: A Robust Cooperative MAC Layer Framework}
\label{WLAN-MAC-R-DSTC}

While R-DSTC has been mainly studied in
the PHY layer, an efficient MAC layer protocol is essential in
order to enable its use in a real environment and to translate its
PHY layer benefits to performance gain in the upper layers. This
section presents a robust MAC layer protocol, called
\emph{Space-Time coding in Cooperative MAC} (\emph{STiCMAC}), in
support of R-DSTC in an IEEE 802.11 WLAN environment. In this
paper, we consider a WLAN operating in the infrastructure mode,
where an access point (AP) works as a central unit. The proposed
MAC protocol is mainly composed of two parts: (1) a three-way
handshake, which includes relay recruiting and acknowledgements;
(2) cooperative two-hop data transmission. The three-way
handshaking takes care of all signaling among the source,
destination and relays. During the handshaking procedure, relays
are recruited simultaneously and opportunistically according to
each relay's instantaneous channel conditions, while no other stations, except for the selected relays, needs to know these channel conditions. Cooperative two-hop data
transmission occurs when relays receive the necessary transmission
parameters. The details of \emph{STiCMAC} are explained in the
rest of this section.

\vspace{-0.15in}
\subsection{Wireless LAN Medium Access Control Overview}
\label{WLAN-DCF}
\vspace{-0.05in}

In the IEEE 802.11 WLAN standard~\cite{80211-2007}, Distributed Coordination Function (DCF) is 
the mandatory MAC protocol. Since DCF is contention based, stations employ carrier sensing multiple
access/collision avoidance (CSMA/CA) algorithm to resolve
collisions. Under this scheme, each station can starts a packet
transmission only if it senses the channel to be free. However, due to sensing range limitations, two stations
could be sending to a common receiver simultaneously. This
phenomena is referred to as the hidden node problem. In order to
avoid such scenarios, virtual carrier sensing is employed, by
means of the Request-to-Send (RTS) and Clear-To-Send (CTS) frames.
These two control frames broadcast the duration for the upcoming
data transmission so that stations that do not participate in this
transmission withhold their own transmissions until the end of the
ongoing packet transmission. In this paper, we focus on the DCF
mode with RTS/CTS messaging and develop a cross-layer framework
for a distributed cooperative system based on IEEE 802.11. STiCMAC with RTS/CTS turned off is presented in the appendix of this paper.

\vspace{-0.15in}
\subsection{Protocol Design for R-DSTC in WLANs}
\label{MAC-R-DSTC}
\vspace{-0.05in}

In this subsection, we introduce the
\emph{STiCMAC} protocol that enables R-DSTC in an
infrastructure-based WLAN under DCF mode. \emph{STiCMAC} enables
relay discovery and concurrent cooperative transmissions from all
relays to the destination. Without loss of generality we consider that the source of a transmission is a
station while the destination is the AP. A symmetric scheme with
the same characteristics can be applied for the downlink
transmission (from the AP to the stations).

\begin{figure}
\centering
   \includegraphics[width=.7\textwidth]{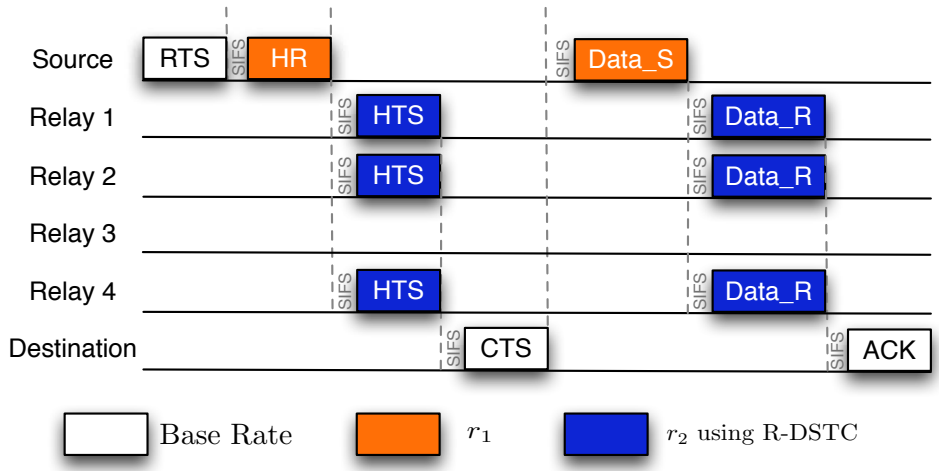}
\caption{\label{fig:RDSTCsignaling} Signaling procedure for
\emph{STiCMAC} based cooperation.}
\end{figure}

In order to enable all relays to forward a packet in unison, the
 MAC layer needs to provide critical parameters for the cooperative
 transmission. The required transmission include the transmission rates for both
 hops, and the underlying space-time code for the second hop. Let us denote $r_1$ as the first-hop rate, $r_2$ as the second-hop rate, and $L$ as the STC dimension.
 We assume that R-DSTC uses a class of underlying orthogonal STC's parameterized by the code dimension $L$. A proper joint selection of these parameters can optimize the MAC layer
 performance. Details of such an optimization will be provided in Section
 \ref{sec:rateadapt}. Additionally, the MAC layer must also provide timing information for both hops, as a data packet undergoes
 a two-hop transmission. In \emph{STiCMAC}, a three-way handshaking procedure is initiated by the source to disseminate these transmission parameters,
  followed by the data transmission. Fig. \ref{fig:RDSTCsignaling} illustrates how \emph{STiCMAC} works for a single packet transmission, which consists of the following steps:

 {\em a. The Three Way Handshaking Phase}

\begin{enumerate}
\item The source station  initiates the handshaking by
transmitting a RTS frame at the base rate in compliance with the
IEEE 802.11 protocol. The RTS frame reserves the channel for
subsequent signaling and data messages. The source continues with
the transmission of the second control frame, called
\emph{Helper-Recruiting} (HR) frame, a \emph{short inter-frame
spacing} (SIFS) period after the transmission of the RTS frame.
This HR frame is transmitted at the chosen first-hop rate $r_{1}$
using a corresponding physical layer modulation level and channel
coding rate. Only those stations that have a channel strong enough
to decode the HR message are likely to receive the subsequent data
packet correctly at the same rate. Thus, all stations receiving
the HR frame correctly are recruited as relays for the current
data packet forwarding. Since recruiting of the relays is
conducted on the fly, this procedure is fully decentralized. The
exact set of recruited relays may vary from packet to packet due
to channel variations or mobility, enabling fully opportunistic
use of relays. The HR frame contains the underlying STC dimension
$L$ and the hop-2 rate $r_{2}$, which is characterized by a
modulation scheme and channel coding rate.

\item A SIFS time after the HR  frame, the recruited relays send
in unison the helper-to-send (HTS) frame using R-DSTC. The
transmission is at the second-hop rate $r_{2}$, using an STC of
size $L$. The HTS message is jointly transmitted by all relays
that successfully decoded the HR message from the source station.
Since a single STC is employed by all the relays, only a
single HTS message is received and decoded by the destination
station without causing a collision. The HTS frame is employed for
the following reasons. Firstly, it is used as an acknowledgement
to the source station that one or more relays have been recruited.
Secondly, the destination station, as long as it receives the HTS
frame correctly, can verify that those relays can indeed support a
rate $r_{2}$ transmission to the destination, even though it doesn't explicitly know which stations act as relays. Thirdly, HTS frame
helps
to alert the hidden terminals around the relays and avoid a possible collision. 

\item The destination responds with a CTS frame, which signals the
end of the three-way handshaking among all participants. The above
handshaking procedure reduces the likelihood of a data packet
collision which is especially in the case of a long data packet.
\end{enumerate}

{\em b. Data Transmission Phase}
\begin{enumerate}

 \item In the data transmission stage, the source
station proceeds with sending the data frame over the first hop,
at rate $r_{1}$. We call this frame \emph{Data-S} frame.

\item The recruited relays cooperatively transmit the data frame
over the second hop, at rate $r_{2}$. We denote this frame as
\emph{Data-R}. The transmission employs an STC dimension of $L$.

\item The destination station finishes the procedure by sending
back to the sender an \emph{Acknowledgement (ACK)} message in
order to confirm that the data packet is successfully received.
\end{enumerate}

The above protocol is backward compatible with standard IEEE 802.11 WLAN protocol since RTS/CTS follows the same format as defined in standard WiFi. Legacy stations can read the \emph{Duration} field and set their \emph{Network Allocation Vector (NAV)}, which indicates how long the surrounding nodes must defer from accessing the channel. Thus legacy stations can co-exist with STiCMAC stations, even though they cannot participate in the cooperative transmissions. The newly introduced \emph{HR} and \emph{HTS} messages do lead to some additional overhead, which is evaluated in Section \ref{PerformanceEvaluation}.

\vspace{-0.2in}
\section{Rate Adaptation} \label{sec:rateadapt}

Rate adaptation refers to the adjustment of the values of the
transmission parameters,  e.g., for \emph{STiCMAC}, the first hop
rate $r_1$, second hop rate $r_2$ and STC dimension $L$, based on
the network conditions. In this section, we develop a rate
adaptation mechanism to maximize the end-to-end user rate while
meeting an acceptable error probability. Our rate adaptation
scheme is subject to an end-to-end packet error rate (PER)
threshold $\gamma$, before MAC layer retransmissions are
initiated. The selection of $\gamma$ affects the system
performance. A high $\gamma$ leads to too many retransmissions due
to high packet loss at the MAC layer, while a low $\gamma$ leads
to an under-utilized bandwidth since the communication link could
support higher modulation and coding rates. Rate adaptation also requires calibration in the physical layer~\cite{hardwareCalibrate} for a practical system. Considering that the main focus of this paper is the MAC protocol design, we do not address this issue in detail and assume that all stations have been calibrated.

We assume each station supports a set of transmission rates $r \in
\{R_0, R_1, \ldots, R_p$\}, where $R_0$ is the base rate at which
the stations exchange control information, i.e., RTS/CTS, and
$R_0<R_1<\cdots<R_p$. A given $r$ is identified by the modulation
level $M_r$ and the channel coding level $C_r$. In addition, we
denote the STC code rate as $R_c$. We assume an additive white
Gaussian noise (AWGN) channel with independent slow Rayleigh
fading between each pair of stations and between the stations and
the AP. Each fading level is assumed to be longer than a packet
duration. All stations have a symbol energy of $E_{s}$ and the
power spectral density of noise signal is $N_0/2$.

In order to present our rate adaptation scheme, we first formulate
the PHY layer error rates, i.e. per-hop bit error rate (BER),
per-hop PER and
end-to-end PER performance. 
Along with R-DSTC, we also calculate for comparison the PHY layer
error rates for the direct transmission scheme, the two-hop
single-relay ({\em CoopMAC}) scheme and the DSTC scheme.

The BER and PER for a direct connection between stations $i$ and
$j$ are respectively denoted as $P_{b}^{ij}(r,h_{ij})$ and
$P_{p}^{ij}(r,h_{ij})$ for a fixed rate $r$ and instantaneous
channel gain $h_{ij}$ between stations $i$ and $j$. Let us denote
$i=s$ when station $i$ is the source and $j=d$ when station $j$ is
the destination. Assuming the instantaneous channel gain vector
from all relays to the destination is denoted as
$\mathbf{h^{(2)}}$, where $\mathbf{h^{(2)}}=[ \dots, h_{jd},
\dots]$, the BER and PER between relays and the destination for
DSTC with space-time codeword dimension $L$ are denoted by
$P_{b,hop2}^{DSTC}(r,L, \mathbf{h^{(2)}})$ and
$P_{p,hop2}^{DSTC}(r,L,\mathbf{h^{(2)}})$, respectively.
Similarly, when the instantaneous channel gain vector
$\mathbf{h^{(2)}}$ weighted by the instantaneous random matrix
$\mathbf{R}$ is defined as $\mathbf{h^{(2)}R}$, the BER and PER
between relays and the destination for R-DSTC with space-time
codeword dimension $L$ are denoted as
$P_{b,hop2}^{R-DSTC}(r,L,\mathbf{h^{(2)},R})$ and
$P_{p,hop2}^{R-DSTC}(r,L,\mathbf{h^{(2)}, R})$, respectively, for
a given $r$. When these instantaneous PHY layer error rates are
averaged over channel fading levels, the average BER and PER rates
can be obtained accordingly. Table \ref{tab:parameters} lists all
parameters and notation used in this section. Based upon the
analysis of PHY layer error rates, rate adaptation optimizes the
transmission parameters for all transmission strategies with an
objective to maximize the end-to-end rate for each station while
ensure an end-to-end PER bounded by $\gamma$.

\vspace{-0.15in}
\subsection{R-DSTC PHY Layer Per-hop BER Performance} \label{sec:berperformance}
\vspace{-0.05in}

In the suggested two-hop framework, note that the BER for the
first hop between the source and a potential relay can be computed
using the direct link formulation. We denote the second hop between
all the relays and the destination  as the {\em cooperative R-DSTC
link}. This subsection formulates the BER performance for both
links.

\subsubsection{BER performance for a direct link}\label{sec:singlelinkber}
In a direct link with an instantaneous channel gain $h_{ij}$ along
with rate $r$ (corresponding to modulation level $M_r$, assuming
square modulation) between stations $i$ and $j$, the symbol error
rate is given by 
\begin{equation}
P_{s}^{ij}(r,h_{ij})=1-[1-P_{\sqrt{M_r}}]^2, \label{eq:ser}
\end{equation}
with
\begin{equation}
P_{\sqrt{M_r}}=2
(1-\frac{1}{\sqrt{M_r}})Q\left({\sqrt{\frac{3E_{s}\|h_{ij}\|^2
}{(M_r-1)N_0}}}\right), \label{eq:prob}
\end{equation}
where $Q(x)=\int_x^\infty \frac{1}{\sqrt{2 \pi}} e^{-z^2 / 2} \ud
z$. If Gray coding is used in the constellation, the approximate BER is
\begin{equation}
P_{b}^{ij}(r,h_{ij})\approx \frac{1}{\log_2M_r}
P_{s}^{ij}(r,h_{ij}). \label{eq:BER_mod}
\end{equation}

When station $i$ is the source and station $j$ is a relay, Eq.
(\ref{eq:BER_mod}) describes the instantaneous BER performance
from the source to a relay, since each relay decodes the source
signal independently. Additionally, Eq. (\ref{eq:BER_mod}) gives
the per-hop BER performance for a \emph{CoopMAC} system
\cite{coopmacliu} for both hops, when  a single relay is employed
without combining the first and second hop signal at the receiver.

\subsubsection{BER performance for the cooperative R-DSTC link}\label{sec:RDSTCber}
Assuming $n$ relays, i.e., stations $1 \dots n$, successfully
decode the source signal, each relay forwards the signal over the
{\em cooperative R-DSTC link}. Suppose the STC used by the relays
has a dimension of $L \times K$ ~\cite{Jafarkhanistcbook}. During
symbol interval $m$, $m=1,2,\cdots,K$, the forwarded signal from
relay $j$ is given by
\begin{equation}
z_j(m)=\sqrt{E_{s}}\mathbf{w}_j\mathbf{X}(m),
\end{equation}
where $j=1,2, \cdots, n$, and $\mathbf{w}_j$ is the random vector
at relay $j$, as described in Section \ref{R-DSTC}. Here,
$\mathbf{X}(m)$ is the $m$th column of the STC. The received
signal at the destination, during the $m$th symbol interval is
given by,
\begin{equation}
y(m)=\mathbf{h^{(2)}}\mathbf{Z}(m)+w(m)=\sqrt{E_{s}}\mathbf{h^{(2)}}\mathbf{R}\mathbf{X}(m)+w(m),
\label{rdstcreceivedsignal}
\end{equation}
where $w(m)$ denotes the AWGN at the $m$th symbol, and
$\mathbf{Z(m)}=[z_1(m) \; z_2(m) \; \ldots \; z_n(m)]^T$. Hence,
the destination station observes a space-time coded signal with
equivalent channel gain vector $\mathbf{h^{(2)}R}$. The
destination only needs to estimate $\mathbf{h^{(2)}R}$ for STC
decoding. By using the orthogonality of the underlying STC
\cite{Jafarkhanistcbook}, the BER of the cooperative R-DSTC link
using rate $r$, $L$ and a fixed $\|\mathbf{h^{(2)}}\mathbf{R}\|$,
denoted by $P_{b, hop2}^{R-DSTC}(r, L, \mathbf{h^{(2)},R})$, can
be computed by replacing $\|h_{ij}\|$ with $\|\mathbf{h^{(2)}R}\|$
in Eq. (\ref{eq:BER_mod}).

Note that for an orthogonal DSTC with a space time codeword
dimension $L$, each of the $L$ relays is assigned an antenna
index. Hence, there is no randomization and
$\mathbf{R}=\mathbf{I}$. Therefore, the BER of DSTC, denoted by
$P_{b, hop2}^{DSTC}(r, L, \mathbf{h^{(2)}})$, can also be
determined from Eq. (\ref{eq:BER_mod}) by replacing $\|h_{ij}\|$
with $\|\mathbf{h^{(2)}}\|$.

\vspace{-0.15in}
\subsection{End-to-End PER Performance for R-DSTC}\label{sec:perfading}
\vspace{-0.05in}

We derive the average PER by simulations, using the BER
formulation in Section \ref{sec:berperformance} along with the
channel code. Each relay adopts a convolutional code $C_r$ for
error correction, for a given rate $r$. In our simulation, for
each hop we first produce the coded bits and then generate random
errors according to the computed instantaneous BER. This bit
stream is then fed into the convolutional decoder to produce the
decoded bit sequence. The instantaneous PER for both hops is then
obtained by comparing the output bit stream at the destination
with the original bit stream.

Let us denote by $\mathcal{RS}$ the instantaneous set of relays
that correctly decode the source signal in the first hop and
jointly forward the signal over the second hop. In the first hop,
the instantaneous channel gain vector is defined as
$\mathbf{h^{(1)}}$ including all $h_{ij}$, $j \in \mathcal{RS}$. Consequently, for a given $r_1$, $r_2$, $L$, the end-to-end
instantaneous PER between station $i$ and the destination for the
R-DSTC scheme is given by
\begin{equation}
\small {P}_{p}^{R-DSTC}(r_1, r_2, L, {\mathbf{h^{(1)}}},
\mathbf{h^{(2)}, R}) = 1- \sum_{\mathcal{RS} \in \mathcal {PS
(S}_i) } P\mathcal{(RS}) \times { \left(
1-P_{p,hop2}^{R-DSTC}{\mathbf(r_2, L, \mathbf{h^{(2)},R}})\right)}
, \label{eq:per_rdstc_inst}
\end{equation}

where
\begin{equation}
P\mathcal{(RS}) = \prod_{j \in \mathcal{RS}} (1-P_{p}^{ij}(r_1,
h_{ij})) \times \prod_{j \notin \mathcal{RS}} P_{p}^{ij}(r_1,
h_{ij}), \label{eq:prob_rs}
\end{equation}

In Eq. (\ref{eq:per_rdstc_inst}), $S_i$ denotes all stations
excluding the source station $i$, and $\mathcal{PS (S}_i)$ is the
power set of $\mathcal{S}_i$. $P\mathcal{(RS})$ is the probability
that an instantaneous relay set $\mathcal{RS}$ is recruited.
$P_{p}^{ij}(r_1, h_{ij})$ denotes the instantaneous PER between
station $i$ and relay $j$ over the first hop, while $P_{p,
hop2}^{R-DSTC}{\mathbf(r_2,L, \mathbf{h^{(2)},R})}$ denotes the
instantaneous PER from all relays in $\mathcal{RS}$ to the
destination, over the second hop.

Following Eq. (\ref{eq:per_rdstc_inst}) for instantaneous
end-to-end PER, the average end-to-end PER can be obtained by
averaging over all first and second hop channel gains and the
random weight vector $\mathbf{R}$, and is given by
\begin{equation}
{P_p^{R-DSTC}(r_1, r_2,L)} =
\mathbb{E}_{\mathbf{h^{(1)},h^{(2)},R}}\left\{{P}_{p}^{R-DSTC}(r_1,
r_2, L, {\mathbf{h^{(1)}}}, \mathbf{h^{(2)}, R})\right\}.
\label{eq:rdstce2ePEP}
\end{equation}

\vspace{-0.15in}
\subsection{End-to-End PER Performance for Other Schemes}\label{sec:end2end}
\vspace{-0.05in}

Below we formulate the end-to-end PER performance for the other
schemes in order to provide comparison with the proposed R-DSTC scheme.
For the direct transmission scheme, the average end-to-end PER
between the source station $i$ and the destination can be found
using the direct-link instantaneous PER, and is given by
\begin{equation}
P_{p}^{direct}(r)=\mathbb{E}_{h_{id}}\left\{P_{p}^{id}(r,
h_{id})\right\},\label{eq:e2eperdirect}
\end{equation} where the direct rate is $r$ and $P_{p}^{id}(r, h_{id})$ denotes the
instantaneous PER between source station $i$ and the destination
for a given $h_{id}$ and $r$.

For \emph{CoopMAC} scheme, the average end-to-end PER performance
of source station $i$ also depends on the chosen single relay $j$,
$j \in \mathcal{S}_i$ and is given by
\begin{equation}
P_{p}^{coop}(r_1, r_2,j)=\mathbb{E}_{h_{ij}, h_{jd}}\left\{
1-\left(1-P_{p}^{ij}(r_1,h_{ij})\right) \times
\left(1-P_{p}^{jd}(r_2,h_{jd})\right)\right\}, \label{eq:percoop}
\end{equation}
where $h_{ij}$ and $h_{jd}$ denote the instantaneous channel gain
for the first and second hops, respectively. $P_{p}^{ij}(r_1,
h_{ij})$ and $P_{p}^{jd} (r_2, h_{jd})$ denote the instantaneous
PER for the two hops, for a given $h_{ij}$ and $h_{jd}$.

For DSTC, the instantaneous end-to-end PER between source $i$ and destination $d$ depends
on a fixed and predefined relay set, denoted as $\mathcal{RS}_i$,
and is formulated as
\begin{equation}
P_p^{DSTC}(r_1, r_2,L, \mathcal{RS}_i) = \mathbb{E}_{
{\mathbf{h^{(1)}}}, \mathbf{h^{(2)}}}\left\{P_{p}^{DSTC}(r_1, r_2,
L, \mathcal{RS}_i,{\mathbf{h^{(1)}}}, \mathbf{h^{(2)}})\right\}.
\label{eq:dstce2ePEP}
\end{equation}

In Eq. (\ref{eq:dstce2ePEP}), ${P}_{p}^{DSTC}(r_1, r_2, L,
\mathcal{RS}_i, {\mathbf{h^{(1)}}}, \mathbf{h^{(2)}}) $ denotes
the instantaneous end-to-end PER and is given by
\begin{equation}
{P}_{p}^{DSTC}(r_1, r_2, L, \mathcal{RS}_i, {\mathbf{h^{(1)}}},
\mathbf{h^{(2)}}) = 1 - \sum_{_{\mathcal{RS}' \in \mathcal {PS
(RS}_i) }} { (1-P_{p, hop2}^{DSTC}{\mathbf(r_2, L,
\mathbf{h^{(2)}})}) \times P({\mathcal{RS}'})},
\label{eq:per_dstc_inst}
\end{equation}
where
\begin{equation}
P({\mathcal{RS}'}) =  \prod_{j \in \mathcal{RS}'}
(1-P_{p}^{ij}(r_1, h_{ij})) \times \prod_{j \notin \mathcal{RS}'}
P_{p}^{ij}(r_1, h_{ij}), \label{eq:per_dstc_subset}
\end{equation}
$\mathcal{RS}'$ is an instantaneous subset of relays from
$\mathcal{RS}_{i}$ that participate in relaying and $P_{p,
hop2}^{DSTC}{\mathbf(r_2, L, \mathbf{h^{(2)}})}$ denotes the
instantaneous PER  between relays and the destination for fixed
rate $r_2$, STC dimension $L$ and channel gains $\mathbf{h^{(2)}}$
based on relays in $\mathcal{RS}'$.

\subsection{Optimizing end-to-end rate and the choice of transmission parameters}
We now describe how to choose the optimal transmission parameters
in order to maximize the end-to-end transmission rate for each
station, while ensuring the end-to-end average PER is bounded by
$\gamma$. Even though our emphasis is on R-DSTC, we also discuss
direct transmission, \emph{CoopMAC} and DSTC schemes. In our rate
adaptation, every scheme relies on knowledge of the average
channel statistics rather than the instantaneous channel gains,
thus making it suitable for a WLAN where the average
channel statistics change slowly. Assuming the WLAN channel is
reciprocal, the source station can estimate the average channel
statistics for direct transmission and relayed transmission by
listening beacons from AP and overhearing transmissions of other
stations. For WLAN systems are typically a stationary
or low-mobility environment, the average channel statistics are
measured and reported every few seconds, producing only negligible
performance loss in throughput. We also discuss rate adaptation based only on number of users in the network.

The set of transmission parameters is different for each scheme
and is discussed below. In this paper, all transmission parameters
are computed at the source station, as opposed to the protocol in
\cite{RcoopMAC} that conducts the computation at the destination
station. The maximum end-to-end rate of station $i$ is achieved by
minimizing the end-to-end transmission time for each scheme. In
addition, we will also discuss the channel information assumed by
each scheme for rate adaptation.

\subsubsection{Direct transmission scheme}
When the source station $i$ transmits to the destination directly,
assuming the source station knows the channel statistics to the
destination, the optimal transmission parameter is the
transmission rate $r$ and the optimum rate $r^*$ is given by
\begin{equation}
r^{*}=\max r \;\; \; \;\;\; s.t. \;\;  P_{p}^{direct}(r) \leq \gamma,
\label{directrateadapt}
\end{equation}
where $P_p^{direct}(r)$ is given by Eq. (\ref{eq:e2eperdirect}).
Note that the optimal transmission rate $r$ is modified whenever
the source or destination move to a new location, since the
average channel gain for the direct link changes.

\subsubsection{Two-hop single-relay (CoopMAC) scheme}
We assume there is no signal combining at the destination. A
practical MAC protocol for this scheme is {\em
CoopMAC}~\cite{coopmacliu}. Assuming the source knows the channel
statistics between itself and all other stations and between other
stations and the destination, the transmission parameters include
$r_1$, $r_2$ and the selected relay $j$. In {\em CoopMAC}
~\cite{coopmacliu}, the optimum relay information is stored in a
{\em CoopTable} at each source station. The optimum rates $r_1^*$,
$r_2^*$ and the best relay $j^*$ are selected by minimizing the
end-to-end transmission time over two hops, and is formulated by
\begin{eqnarray}
(r_{1}^{*}, r_{2}^{*}, j^{*})=  \underset{r_{1}, r_{2}, j}{\operatorname{arg\,min}}  \; \frac{1}{r_{1}}+\frac{1}{r_{2}}  \;\;\;\;\;\; s.t. \;\; P_{p}^{coop}(r_1, r_2, j) \leq \gamma,
\label{cooprateadapt}
\end{eqnarray}where ${P_{p}^{coop}(r_1, r_2, j)}$ is given by Eq.(\ref{eq:percoop}).  When the
network topology changes, e.g., the source, destination or any
other station move to new locations, the optimal parameters are
reselected using Eq. (\ref{cooprateadapt}). Hence, {\em CoopMAC}
is more suitable for a stationary environment with low mobility.
When the stations move rapidly, the demand for collecting the
global channel knowledge leads to a large overhead for the system. An
inaccurate estimation of the channel results in a non-optimal rate
adaptation scheme and thus degrades the system performance, as further illustrated in Section~\ref{PerformanceEvaluation}.

\subsubsection{DSTC scheme}
Assuming the available space-time codewords have dimensions
denoted by $T = \{L_1,L_2,\dots, L_{max}$\}, DSTC needs to select
its relay set $\mathcal{RS}_i$ consisting of $L$ relays, where $L
\in T$ and $L= \left|{\mathcal{RS}_i}\right|$. Thus, the
transmission parameters are rates $r_1$, $r_2$, $L$ and
$\mathcal{RS}_i$. Similar to \emph{CoopMAC}, the source station is
assumed to know the average channel statistics between itself and
each stations and between other stations and the destination. The
optimum transmission parameters can be obtained by,
\begin{eqnarray}
(r_{1}^{*}, r_{2}^{*}, L^{*}, \mathcal{RS}_i^*)=  \underset{r_{1}, r_{2}, L, \mathcal{RS}}{\operatorname{arg\,min}}  \; \frac{1}{r_{1}}+\frac{1}{R_{c} r_{2}} \;\;\;\;\;\; s.t. \;\; {P_{p}^{DSTC}(r_1, r_2, L, \mathcal{RS}_i)} \leq \gamma,
\label{dstcrateadapt}
\end{eqnarray} where ${P_{p}^{DSTC}(r_1, r_2, L,
\mathcal{RS}_i)}$ is given by Eq.(\ref{eq:dstce2ePEP}), and
${R_c}$ is the rate of the orthogonal STC with dimension $L$. It
is known that it is only possible to have full rate ($R_c$ = 1)
orthogonal STC for $L$ = 2, otherwise $R_c < 1$
\cite{Jafarkhanistcbook}. With $N$ stations in the single-cell
WLAN excluding the destination AP, there are $\sum_ {L \in
T}\binom{N-1}{L}$ possible relay sets $\mathcal{RS}_i$ containing $L$ relays. An
exhaustive search for all possible relays in $\mathcal{RS}_i$
leads to a combinatorial complexity, it is prohibitively expensive
to solve online.

In order to reduce the complexity of relay selection, we propose a greedy algorithm and use it to evaluate DSTC performance in Section \ref{PerformanceEvaluation}. The basic idea is to sequentially add $L$ relays to the optimal relay set. For each step, we find a single relay that, when combined with the relays selected in the previous steps, will maximize the end-to-end throughput if DSTC is used to assist transmissions from the source. For the first relay, we choose the best relay from the $N$-1 stations that maximizes the end-to-end rate in a two-hop manner (single relay based CoopMAC~\cite{coopmacliu}) and add it into the relay set. Then, the second relay is chosen from the remaining $N$-2 stations in such a way that it can achieve the maximal end-to-end rate along with the first selected relay, using DSTC. Such a selection is iterated until all $L$ relays are picked and added to the relay set. Our simulation shows only 5\% throughput difference between this greedy algorithm and exhaustive search when $L$=2 and $N$=10.


Like {\em CoopMAC}, DSTC need to reselect the transmission parameters  whenever the network topology changes and incurs a large amount of channel estimation overhead especially in a mobile environment.

\subsubsection{{STiCMAC} scheme}
One difficulty of the \emph{CoopMAC} and \emph{DSTC} strategies is
in choosing the optimal $L^*$ ($L^*=1$ for \emph{CoopMAC} and
$L^*>1$ for \emph{DSTC}) relays out of the $N-1$ other stations. In contrast, the R-DSTC based
\emph{STiCMAC} eliminates such a requirement, and thus the
transmission parameters only include rates $r_1$, $r_2$, and the
STC size $L$, and not the relay set. For
the  \emph{STiCMAC} strategy, we develop two classes of rate
adaptation algorithms based upon different channel knowledge:

\begin{itemize}
    \item The  \emph{STiCMAC-CS} scheme is assumed to have the same channel knowledge of channel statistics as
   in DSTC and {\em CoopMAC}. That is, the source station $i$ is assumed to know the
    channel statistics between itself and other stations and
    between other stations and the destination. The optimal transmission
    parameters $r_{1}^{*}, r_{2}^{*}, L^{*}$ are given by
\begin{eqnarray}
(r_{1}^{*}, r_{2}^{*}, L^{*})=  \underset{r_{1}, r_{2}, L}{\operatorname{arg\,min}}  \; \frac{1}{r_{1}}+\frac{1}{R_{c} r_{2}} \;\;\;\;\;\;  s.t. \;\; {P_{p}^{R-DSTC}(r_1, r_2, L)}  \leq \gamma,
\label{rdstcrateadapt_channel_statistics}
\end{eqnarray}
where the end-to-end PEP for R-DSTC is given by Eq.
(\ref{eq:rdstce2ePEP}) and $R_c$ is the rate of the orthogonal STC
of dimension $L$. The rate adaptation scheme for \emph{STiCMAC-CS}
is described in Algorithm. \ref{channelstatisticssearch}. The
optimal set ($r_1^*$, $r_2^*$, $L^*$) is exhaustively searched
over all possible combinations. Each source station executes this
algorithm to find the optimum transmission parameters whenever any
of channel gains change. Similar to the limitation of {\em
CoopMAC} and DSTC, \emph{STiCMAC-CS} requires a global channel
knowledge and thus is relatively costly in a mobile scenario.

\item \emph{STiCMAC-UC} scheme provides rate adaptation with
minimal channel information. Unlike \emph{STiCMAC-CS}, {\em
CoopMAC} and DSTC, the source station $i$ is only assumed to know
the channel statistics between itself and the destination,
together with $N$, the number of stations in the WLAN.
\footnote{Alternatively, a reasonable assumption is for the source
to estimate  the statistics of the relays' links towards itself,
while being unaware of relays-destination average channel
qualities. The performance of such a scheme would be between {\em
STiCMAC-CS} and {\em STiCMAC-UC}.} \emph{STiCMAC-UC} determines
its optimal rate parameters by simply ensuring the average PER
over all possible spatial locations of stations, is bounded by
$\gamma$, assuming all stations are uniformly located using a
random spatial distribution $\chi$, as shown in the following
equation,
\begin{eqnarray}
(r_{1}^{*}, r_{2}^{*}, L^{*})=  \underset{r_{1}, r_{2}, L}{\operatorname{arg\,min}}  \; \frac{1}{r_{1}}+\frac{1}{R_{c} r_{2}} \;\;\;\;\;\; s.t. \;\; \mathbb{E}_{\chi}{\left (P_{p}^{R-DSTC}(r_1,
r_2,L)\right)} \leq \gamma. \label{rdstcrateadapt_user_count}
\end{eqnarray}

\emph{STiCMAC-UC} scheme is described in detail in Algorithm
\ref{usercountsearch} and only depends on the number of stations
in the WLAN without the need for their specific locations. Since
\emph{STiCMAC-UC} requires less information, for a specific
location of users, it yields suboptimal operating parameters
compared to \emph{STiCMAC-CS}. However, \emph{STiCMAC-UC}
eliminates extra signaling for channel measurements and is
suitable for a mobile environment where collecting global channel
statistics is hard and costly. Note that DSTC and {\em CoopMAC}
need to pre-determine the relays before a transmission can be
initiated, hence cannot be based on merely the number of users.
\end{itemize}

For all rate adaptation schemes with full channel statistics, namely \emph{direct}, \emph{CoopMAC}, \emph{DSTC} and \emph{STiCMAC-CS}, we assume optimal parameters are recomputed whenever the channel statistics change. For \emph{STiCMAC-UC}, a two-dimensional look-up table can be
pre-computed for the optimal transmission parameters and saved at
each source, corresponding to the total number of stations, $N$,
and the distance from the source to the AP in each cell. Once a
station enters or leaves the WLAN cell, the BS will broadcast such
user count information in its beacon frame to all stations and
each station can update its optimal transmission parameters. Thus,
\emph{STiCMAC-UC} does not need real-time computation during
network operation. Obviously, in all the relay-assisted schemes,
if the end-to-end rate derived by the used rate adaptation scheme
is lower than the direct transmission rate, the source station
chooses the direct transmission mode instead of cooperation.

%

\vspace{-0.2in}
\section{Performance Evaluation}
\label{PerformanceEvaluation} 

In order to evaluate the performance of the proposed {\em STiCMAC} scheme, we developed a detailed simulation model using OPNET modeler. We compare {\em STiCMAC} with direct transmission, {\em CoopMAC} and DSTC MAC for both stationary and mobile environments. Additionally, all schemes use the rate adaptation algorithm described in Section \ref{sec:rateadapt}. The comparison and evaluation was done on a typical single-cell WLAN.

\vspace{-0.15in}
\subsection{Network Topology and Configuration} \label{networktopology} 
\vspace{-0.05in}

We assume that the considered wireless LAN complies with the IEEE 802.11g
standard and the cell radius is 100 meters. Independent Rayleigh slow
fading among each pair of stations and additive white Gaussian noise is adopted as the channel
model. The simulated system consists of one AP at the center of a
cell and $N$ mobile stations. According to
\cite{Jafarkhanistcbook}, both for DSTC and R-DSTC, a full-rate
orthogonal STC is employed for $L=2$ with $R_c$ = 1, while a $R_c$
= 3/4 rate orthogonal STC is employed for $L=3, 4$. Each AP or
mobile station is equipped with a single omnidirectional antenna.
Our simulations are conducted on the uplink from the mobile
stations to the AP, with the parameters shown in Table.
\ref{tab:WLANConfig}. The simulation results display 90\%
confidence intervals.

\vspace{-0.15in}
\subsection{Mobility Model} \label{mobilitymodel} 
\vspace{-0.05in}

Our simulations are performed for both
stationary and mobile scenarios. In the stationary scenario, all
stations are uniformly distributed within the cell coverage. In the mobile scenario, the stations are assumed to move across
the cell using the \emph{random walk with reflection}
(\emph{RWkRlc}) model \cite{McGuire}. The \emph{RWkRlc} model is
widely adopted to characterize the movement of
mobile stations. The \emph{RWkRlc} model initially deploys
stations randomly according to a uniform distribution over the
cell. Then, it assigns a random speed to each station that is
uniformly distributed in the range [${V_{min}}$, ${V_{max}}$].
Each station picks a random travel duration uniformly distributed
in the range [$T_{min}$, $T_{max}$] and a uniformly distributed random direction. Once a station has walked
for the selected duration of time, it may dwell for a random
amount of time $T_d$ based upon a uniform distribution before it
reselects a new travel duration, speed and direction. In
contrast to the classic \emph{Random Walk} model
\cite{mobilitymodel}, the \emph{RWkRlc}-governed model includes
\emph{reflection} as an additional feature. Namely, whenever a
station reaches the cell boundary during its walk, it will be
reflected by the boundary in a similar way that a ray of light
reflects off a mirror. This reflection functionality will ensure
that the random walk is bounded within a given   cell coverage.
Accordingly,  the \emph{RWkRlc} model produces a uniform spatial
distribution of all stations across the cell and thus enables us
to make a fair comparison with the static scenario. The typical parameters of the \emph{RWkRlc} model we used
are shown in Table \ref{tab:WLANConfig}.

\vspace{-0.15in}
\subsection{Simulation Results} \label{simulationresult}
\vspace{-0.05in}

\begin{figure}[h!]
\centering
\includegraphics[width=.8 \textwidth]{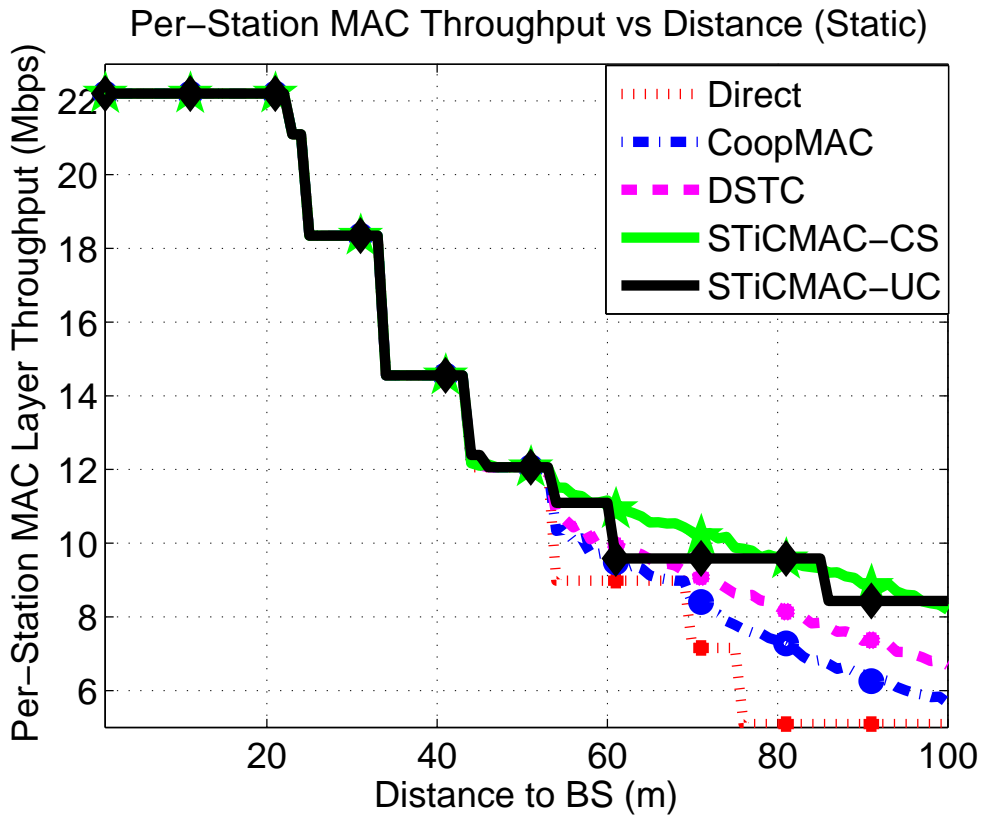}
\caption{Per station MAC layer throughput vs distance (meters) to
AP}
 \label{fig:macratevsdistance}
\end{figure}

Fig. \ref{fig:macratevsdistance} depicts the MAC layer throughput
performance of a single station as a function of its distance to
the AP, assuming $N$=48 stations are uniformly distributed in a
static wireless LAN. When the station is close to the AP, all
schemes fall back to direct transmission  and thus achieve the
same throughput. As the distance to the AP grows, all
the two-hop schemes outperform the direct transmission, since two
high-speed hops provide a higher end-to-end throughput than a
low-speed direct transmission, especially as the stations get
closer to the cell edge. For large distances, \emph{STiCMAC-CS}
and \emph{STiCMAC-UC} schemes show the highest per-station
throughput gains, followed by the \emph{DSTC} and \emph{CoopMAC}.

\begin{figure}[h!]
\centering
   \includegraphics[width=.8\textwidth]{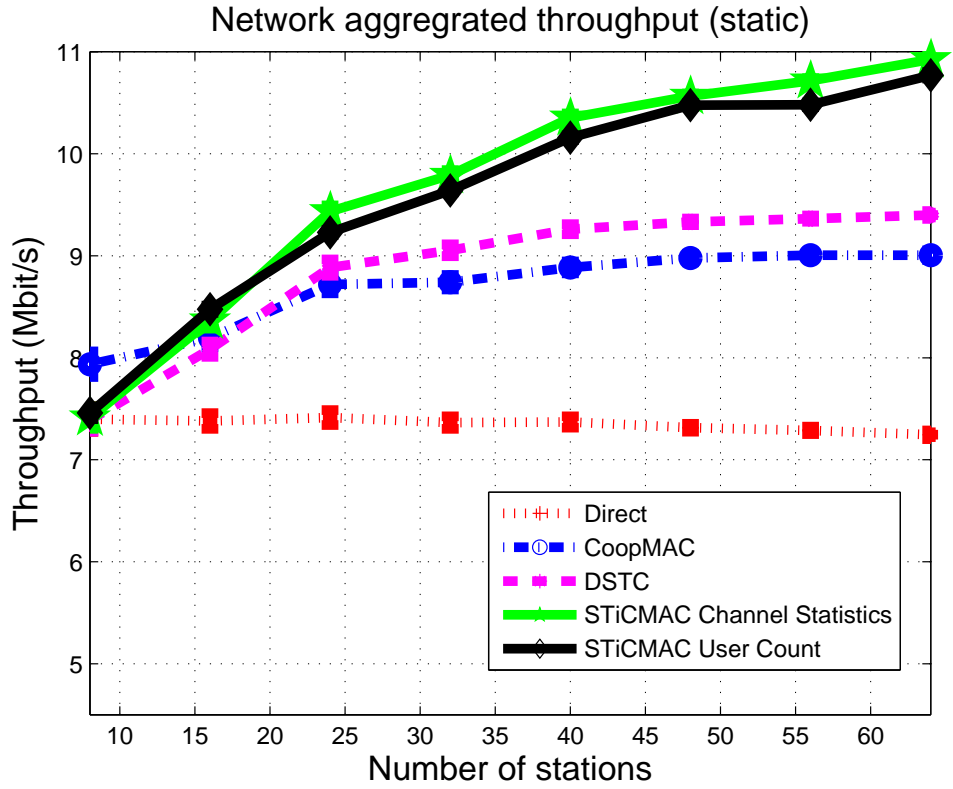}
\caption{Throughput comparison for the static environment.}
\label{fig:staticth}
\end{figure}

Fig. \ref{fig:staticth} displays the comparison of the aggregate
throughput in a stationary environment as a function of $N$, the
number of stations. When the number of stations is less than 16,
the two \emph{STiCMAC} schemes, \emph{STiCMAC-CS} and
\emph{STiCMAC-UC}, provide throughput performance comparable to
\emph{CoopMAC} and {\em DSTC}, while all the cooperative schemes
provide a higher throughput than direct transmission. Note that
for a small number of stations, {\em DSTC} performs worse than the
other two-hop schemes, due to the increased overhead for relay
recruitment. Compared to {\em CoopMAC}, the extra overhead needed
by {\em DSTC} includes the pilot tones (1 time slot for each pilot
which is 9 $\mu$seconds) and relay indices (1 byte for each relay)
sent by the source to the selected relays, as well as the
acknowledgements (1 time slot which is 9 $\mu$seconds for each
relay) from all these relays before every packet transmission is
initiated, as is described in \cite{Jakllari06}. The more relays
are recruited by {\em DSTC}, the higher the overhead. As the
number of stations increases, \emph{STiCMAC} shows a significant
throughput gain over the other schemes (up to 50\% gain over direct) due to the following
reasons: a) A large number of stations lead to a larger
probability of finding more relays, which results in higher
diversity and power gain over the second hop. b) Compared to the
DSTC MAC \cite{Jakllari06}, \emph{STiCMAC} needs substantially
reduced signaling overhead and handshaking. Also, the
overhead of {\em STiCMAC} is constant and independent of the
number of relays, while the DSTC overhead increases as the number
of relays increases. It is also noted that the aggregate throughput of \emph{STiCMAC-UC} is only slightly lower than \emph{STiCMAC-CS}. This is because a sufficiently large number of stations supplies enough potential relays and thus eliminates the need for a
global knowledge of node locations. This validates our argument that {\em STiCMAC} operates
efficiently without a global knowledge of channel statistics.

\begin{figure}[h!]
    \centering
    \includegraphics[width=.8\textwidth]{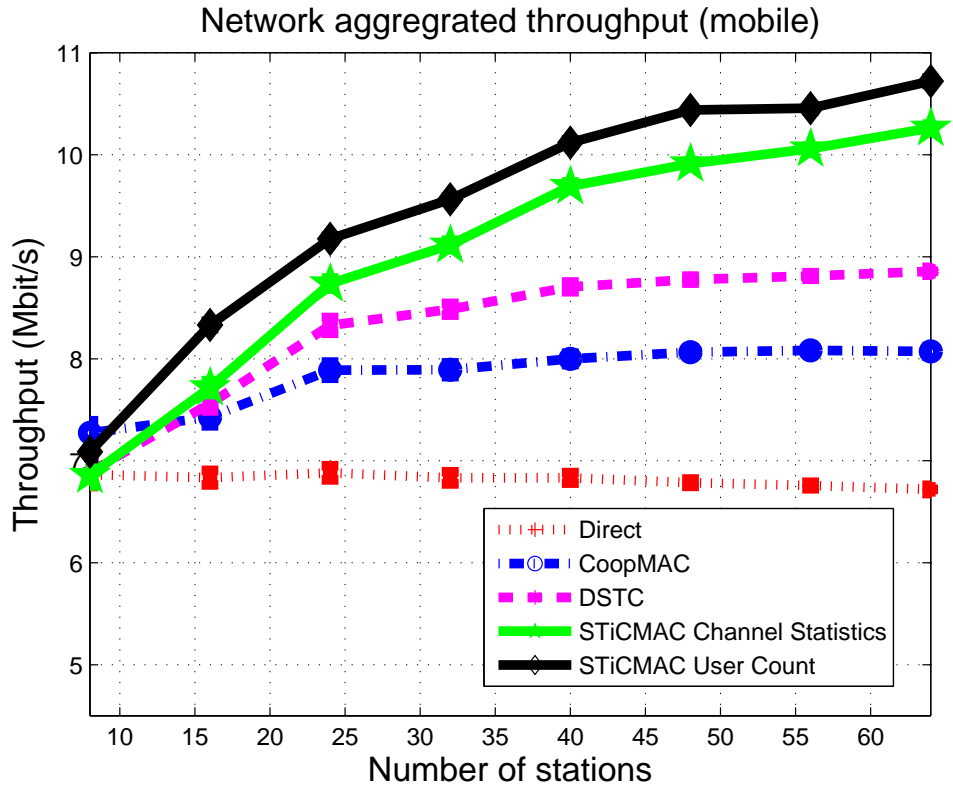}
    \caption{Throughput comparison for the mobile environment.}
    \label{fig:mobileth}
\end{figure}

Fig. \ref{fig:mobileth} depicts the throughput performance of all
schemes in a mobile environment where each station moves according
to the \emph{RWkRlc} model. Under mobility, we assume channel statistics are updated every 2 seconds. Hence each source
station can only perform rate adaptation with 2 second intervals. In contrast to the stationary
scenario, the throughput of all schemes except \emph{STiCMAC-UC}
degrade relative to the static case as mobility leads to an
inaccurate estimation of channel information, resulting in
sub-optimal rate adaptation. For example, in \emph{CoopMAC} and
\emph{DSTC}, the selected relay stations may move away due to
mobility and become unavailable in the forwarding phase. From Fig.
\ref{fig:mobileth}, it is clear that \emph{STiCMAC} schemes
outperform the others in terms of throughput. Under
mobility, \emph{STiCMAC-UC} performance is superior to that of
\emph{STiCMAC-CS}. Therefore in a mobile environment,
\emph{STiCMAC-UC} scheme is preferable since it does not rely on
the instantaneous spatial distribution of all stations for rate
adaptation, and thus leads to more robust throughput performance.

\begin{figure}[h!]
    \centering{\epsfxsize=.8\textwidth \epsfbox{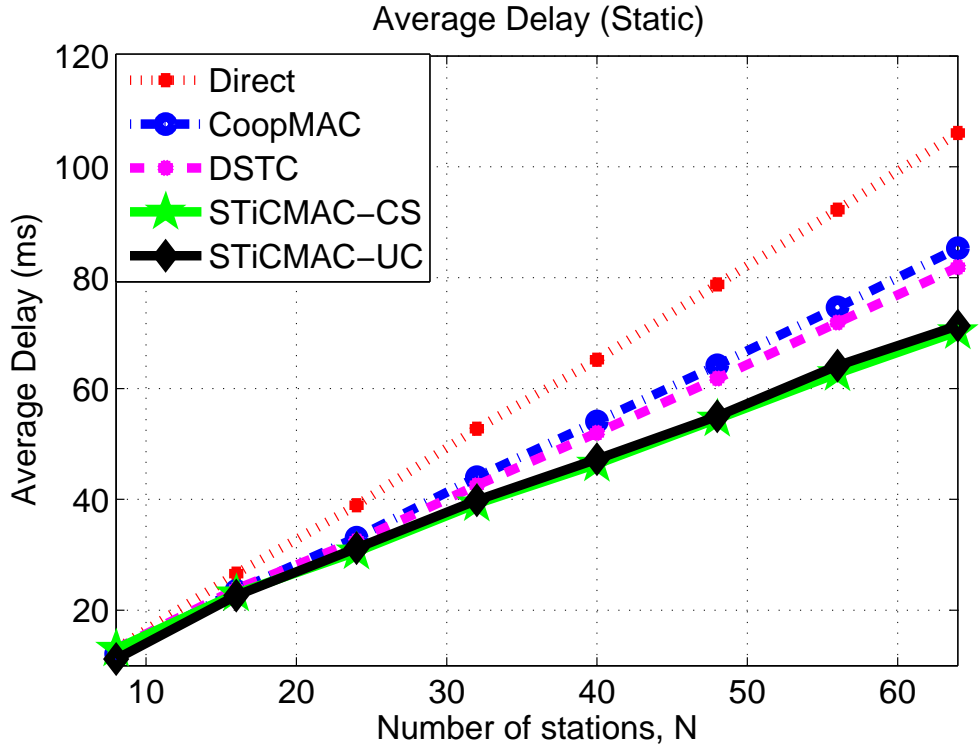}}
    \caption{Medium access delay in a static environment.}
    \label{fig:delaystatic}
\end{figure}

\begin{figure}[h!]
\centering
   \includegraphics[width=.8\textwidth]{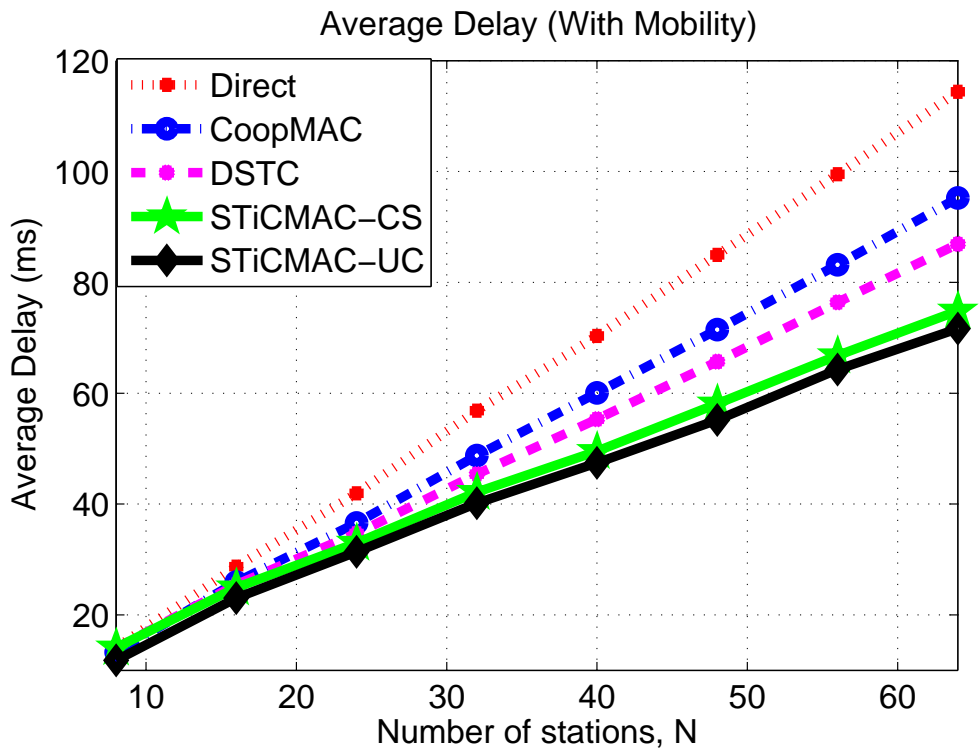}
\caption{\label{fig:delaymobile} Medium access delay in a mobile
environment.}
\end{figure}

Fig. \ref{fig:delaystatic} and Fig. \ref{fig:delaymobile}
demonstrate the medium access delay  for a stationary and mobile
environment respectively under full load. This delay is measured
from the moment that a packet becomes the {head-of-line} packet in
the MAC transmission buffer to the moment that that packet is
successfully received at the MAC layer of the receiver. The
figures reveal that a large number of stations leads to an
increase in medium access delay for all schemes due to the
increased delay before successful access to the channel. However,
\emph{STiCMAC} achieves the lowest delay compared to direct
transmission, \emph{CoopMAC}  and the \emph{DSTC}, since R-DSTC
supports a higher end-to-end rate for each connection, and
therefore decreases the end-to-end transmission time.

\begin{figure}[h!]
\centering
   \includegraphics[width=.8\textwidth]{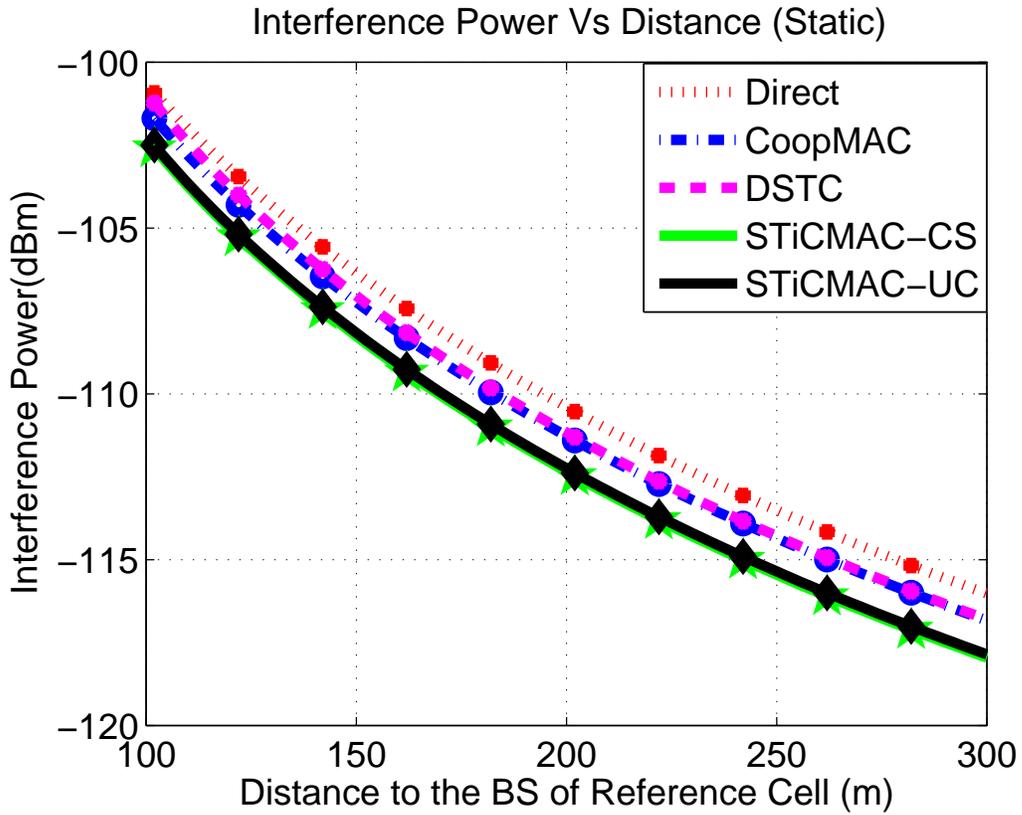}
\caption{\label{fig:powerinterference} Interference power vs
distance (meters) to reference cell.}
\end{figure}

In addition to throughput and delay performance, \emph{STiCMAC}
also reduces the interference generated to neighboring cells when
loaded with traffic at the same level. This is because \emph{STiCMAC} supports a
higher average data rate per packet transmission and thus needs
reduced air time to deliver the same amount of data on an
end-to-end basis, as compared to the other schemes. Consequently,
the average transmission power emanating from the reference cell
is reduced, even though more relays have been recruited. Fig.
\ref{fig:powerinterference} shows the interference in a mobile
scenario where the average interference power generated by a cell
is calculated assuming N=24 users in each cell. The average
interference power is illustrated in
Fig.~\ref{fig:powerinterference} and measured in units of dBm at a
distance of (100 - 300 m) away from the AP of the reference cell.
We observe that both \emph{STiCMAC} schemes generate less
interference compared to {\em DSTC}, \emph{CoopMAC} and direct
transmission. In conclusion, \emph{STiCMAC} generates less
interference at the same MAC layer traffic load compared to the
other schemes.

\vspace{-0.2in}
\section{Conclusion}
\label{Conclusion}

In this paper, we develop a PHY/MAC cross-layer protocol we call
\emph{STiCMAC} by employing R-DSTC in a WLAN system. The
\emph{STiCMAC} protocol incorporates randomized cooperative PHY
layer into the operation of the mandatory DCF MAC of a WLAN to
provide robust cooperative communications using multiple relays. The proposed protocol
is simple and it realizes a significant performance gains in terms of throughput,
delay and interference reduction over various other single-hop and
multi-hop mechanisms (e.g., \emph{CoopMAC} and \emph{DSTC}).
The new MAC is backward compatible with IEEE 802.11. Although only the infrastructure mode is discussed in this paper,
similar ideas also apply to ad hoc WLANs. Compared to previously
known two-hop schemes \cite{coopmacliu, Jakllari06}, STiCMAC
enables a fully \emph{distributed} yet \emph{robust} cooperation
using multiple relays. The signaling and channel feedback overhead
is reduced due to randomized cooperation, resulting in a
significant MAC layer throughput improvement. The robustness of
\emph{STiCMAC} translates to high gains, even in the more
challenging mobile environment. Indeed, the relative gains are
higher for \emph{STiCMAC} in the mobile environment.

\newpage
\begin{appendix}
The operation of IEEE 802.11 MAC is based on carrier sensing, and it also employs virtual carrier sensing (RTS/CTS) to minimize the hidden terminal problem. In order to keep backward compatibility with the legacy system, the design of STiCMAC works with RTS/CTS. 

STiCMAC also works without using RTS/CTS frames. In such a scenario, the source can directly starts a data packet transmission and embed transmission parameters necessary for the second hop transmission (using R-DSTC) in a separate shim header field of the first hop data packet. Any relays that decode it would be able to forward the packet to the destination using R-DSTC.

\begin{figure}[h!]
\centering
   \includegraphics[width=.8\textwidth]{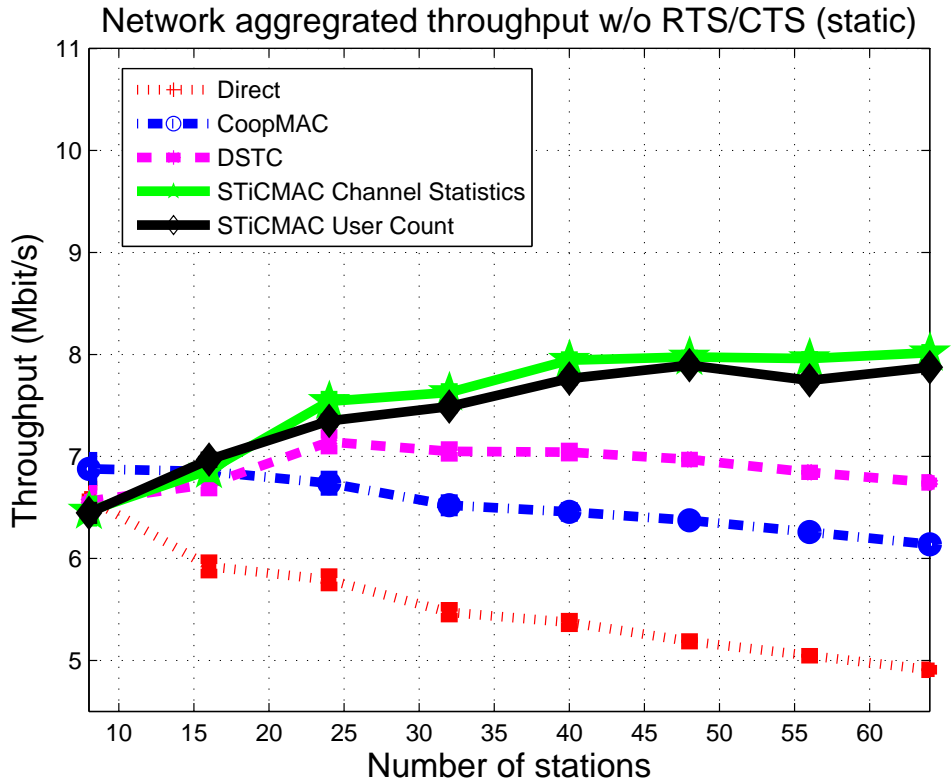}
\caption{\label{fig:thnorts} Throughput comparison without RTS/CTS.}
\end{figure}

We conducted simulations to show how STiCMAC performs without RTS/CTS protection, and the results demonstrate that STiCMAC still outperforms other transmission schemes. The following figures illustrate the throughput and delay without using the RTS/CTS mechanism. It is shown in Fig.~\ref{fig:thnorts} that all schemes display degraded performance as compared to the performance using RTS/CTS. This is because our simulations assume all stations are heavily loaded to show saturated throughput. Therefore, a large number of packet collisions occur because of CSMA/CA based channel access. Additionally, the packet size of the simulation is 1500 bytes, and thus the system performance degrades when not protected by RTS/CTS. While RTS/CTS could be less efficient when the traffic load is light,  STiMAC continues to work well without RTS/CTS. 

\begin{figure}[h!]
\centering
   \includegraphics[width=.8\textwidth]{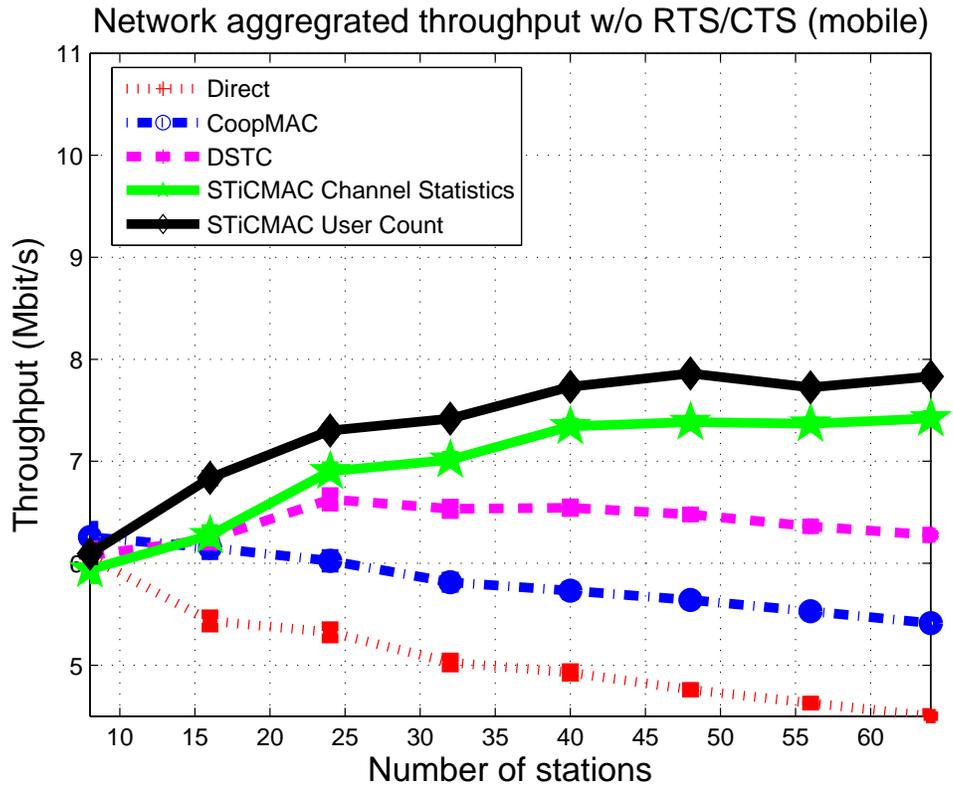}
\caption{\label{fig:thnortsmobile} Throughput comparison for mobile network without RTS/CTS.}
\end{figure}

For a network with moderate mobility (1-2 meters/second), as shown in in Fig.~\ref{fig:thnortsmobile}, the throughput of all schemes is affected. However, STiCMAC is still superior to all other schemes including direct transmission, CoopMAC and DSTC. 
\end{appendix}

\bibliographystyle{IEEEtran}
\bibliography{myreference,phyreference}

\pagebreak


\pagebreak

\pagebreak
\begin{table}[h!] \small
\begin{minipage}[b]{1\linewidth}
 \caption{Notation Used in the Paper}
\vspace{.01in} \centering
\begin{tabular}{c||l} \hline
\label{tab:parameters} \textbf{Notation} & \textbf{Description}
\\ \hline \hline $N$ & Number of stations in a WLAN excluding the AP \\ \hline
$r_1$, $r_2$  & The first and second hop rates\\
\hline $L$ & STC dimension for the underlying space-time code \\
\hline
$\gamma$  & End-to-end PER threshold\\
\hline
$M_r$, $C_r$ & Modulation, channel coding for rate $r$\\
\hline $E_s$ & Symbol energy\\ \hline $N_0$/2 & Power spectral
density of AWGN \\ \hline $h_{ij}$ & Instantaneous channel gain
between stations $i$ and $j$
\\  \hline
 ${\mathbf{h^{(1)}}}$ & Instantaneous channel gain vector between  source and relays\\ \hline $\mathbf{h^{(2)}}$ &
 Instantaneous channel gain vector between relays and destination \\
\hline$\mathcal{RS}_i$ & Deterministic relay set of source station
$i$ for DSTC \\ \hline $\mathcal{RS}$ & Instantaneous relay set of
source station $i$ for R-DSTC \\ \hline $P_{b}^{ij}(r,h_{ij})$ &
BER for a direct connection between stations
$i$ and $j$ for given $r$ and $h_{ij}$. \\
$P_{p}^{ij}(r,h_{ij})$ & PER for a direct connection between stations $i$ and $j$ for given $r$ and $h_{ij}$. \\
\hline $P_{b, hop2}^{R-DSTC}(r, L, \mathbf{h^{(2)},R})$ & BER for
R-DSTC between
relays and destination for given $r$, $L$, $\mathbf{h^{(2)}}$ and $\mathbf{R}$. \\
 $P_{p,hop2}^{R-DSTC}(r, L, \mathbf{h^{(2)},R})$ & PER for  R-DSTC  between relays and destination for given $r$, $L$, $\mathbf{h^{(2)}}$ and $\mathbf{R}$. \\
\hline  $P_{b,hop2}^{DSTC}(r,L, \mathbf{h^{(2)}})$ & BER for DSTC
between relays
and destination for given $r$, $L$ and $\mathbf{h^{(2)}}$. \\
 $P_{p,hop2}^{DSTC}(r, L, \mathbf{h^{(2)}})$ & PER for DSTC between relays and destination for given $r$, $L$ and $\mathbf{h^{(2)}}$. \\
\hline
$P_{p}^{id}(r, h_{id})$ & End-to-end PER for a direct transmission for given $r$ and $h_{id}$. \\
\hline
 $P_{p}^{DSTC}(r_1, r_2,L, \mathcal{RS}_i, {\mathbf{h^{(1)}}}, \mathbf{h^{(2)}})$ & End-to-end PER for DSTC for given $r_1$, $r_2$, $L$, $\mathcal{RS}_i$, ${\mathbf{h^{(1)}}}$ and $\mathbf{h^{(2)}}$. \\
\hline
 $P_{p}^{R-DSTC}(r_1, r_2,L, {\mathbf{h^{(1)}}},\mathbf{h^{(2)},R})$ & End-to-end PER for R-DSTC for given $r_1$, $r_2$, $L$, ${\mathbf{h^{(1)}}}$, $\mathbf{h^{(2)}}$ and $\mathbf{R}$. \\
\hline
$P_{p}^{direct}(r)$ & Average end-to-end PER for a direct transmission for given $r$. \\
\hline $P_{p}^{coop}(r_1, r_2,j)$ & Average end-to-end PER for
CoopMAC for given $r_2$, $r_2$ and relay $j$. \\ \hline
$P_{p}^{R-DSTC}(r_1, r_2,L)$ & Average end-to-end PER for R-DSTC for given $r_1$, $r_2$ and $L$. \\
\hline
 $P_{p}^{DSTC}(r_1, r_2,L, \mathcal{RS}_i)$ & Average end-to-end PER for DSTC for given $r_1$, $r_2$, $L$ and $\mathcal{RS}_i$. \\
\hline

\end{tabular}
\end{minipage} \\
\end{table}

\begin{table}[h!]\small
\begin{minipage}[b]{1\linewidth}
 \caption{Simulation Configuration and Mobility Modeling}
\vspace{.01in} \centering
\begin{tabular}{l||l} \hline
\label{tab:WLANConfig} \textbf{Parameters} & \textbf{Value}
\\ \hline \hline Received $E_s/N_0$ at edge & 1.4 \\ \hline
Path loss exponent & 3.0\\ \hline Propagation Model & ITU-T Indoor Model and Rayleigh fading \\
\hline Spectrum bandwidth& 20 MHz \\
\hline
PHY layer data rates, $r$ & 6, 9, 12, 18, 24, 36, 48, 54 Mbps\\
\hline
Modulation, $M_r$ & BPSK, QPSK, 16-QAM, 64-QAM\\
\hline Channel coding, $C_r$ & Convolutional 1/2, 2/3, 3/4 \cite{80211-2007}\\
\hline Acceptable MAC Layer PER $\gamma$ & 5\%\\
\hline MAC Layer PDU size& 1500 bytes \\
\hline
Contention window size& 0 - 1023 \\
\hline Underlying orthogonal STC dimension, $L$, & 2,3,4 \\
\hline Achievable STC code rates, $R_{c}$ & 1 ($L=2$), 3/4
($L=3,4$)
\\ \hline \hline Min Speed ($V_{min}$) & 1 meter/second \\ \hline
Max Speed ($V_{max}$) & 2 meter/second\\ \hline  Dwell Time during Walk ($T_d$) & 1 second \\
\hline
Min Travel Duration per Step ($T_{min}$) &  2 second\\
\hline
Max Travel Duration per Step ($T_{max}$) &  5 second \\
\hline
\end{tabular}
\end{minipage} \\
\end{table}

\newpage
\begin{algorithm}
\begin{minipage}{1\linewidth}
\caption{Rate Adaptation for STiCMAC Channel Statistics}          
\begin{algorithmic}[1]
 \label{channelstatisticssearch}
\small \STATE The available rate set for both the
first hop ($r_1$) and the second hop ($r_2$) is $\{R_1,R_2,...,R_P\}$, and the set of available orthogonal STC
dimensions for R-DSTC is $L$, where $L  \in
\{L_1,L_2,...,L_{max}\}$. {Initialize $R^* = 0$}.

\FOR{Each possible set of transmission parameters \{$r_1$, $r_2$,
$L \}$}
 \STATE {Find $P_{p}^{R-DSTC}(r_1,
r_2, L)$ for R-DSTC using Eq.(\ref{eq:rdstce2ePEP}}).

\IF{$P_{p}^{R-DSTC}(r_1, r_2, L) < \gamma$ and $\frac {1} {1/r_1 +
1/r_2} > R^*$} \STATE $R^* \leftarrow \frac {1} {1/r_1 + 1/r_2}$,
$L^* \leftarrow L$, $r_1^* \leftarrow r_1$, $r_2^* \leftarrow
r_2$\ \ENDIF

\ENDFOR
\end{algorithmic}
\end{minipage} \\
\end{algorithm}

\begin{algorithm}
\begin{minipage}{1\linewidth}
\caption{Rate Adaptation for STiCMAC User Count}
 \label{usercountsearch}
\begin{algorithmic}[1]
\small \STATE The available rate set for both the
first hop ($r_1$) and the second hop ($r_2$) is $\{R_1,R_2,...,R_P\}$, and the set of available orthogonal STC
dimensions for R-DSTC is $L$, where $L  \in
\{L_1,L_2,...,L_{max}\}$. Suppose all stations are located in the
WLAN cell based on a random distribution function $\chi$.
{Initialize $R^* = 0$}.

 \FOR{Each
possible set of transmission parameters \{$r_1$, $r_2$, $L $\}}
 \FOR{All possible locations of other
stations} \STATE {Find $P_{p}^{R-DSTC}(r_1, r_2, L)$ for R-DSTC
transmission using Eq. (\ref{eq:rdstce2ePEP}}) and average over
all these locations using $\chi$. \ENDFOR \IF{$
\mathbb{E}_{\chi}{\left (P_{p}^{R-DSTC}(r_1, r_2, L)\right)} <
\gamma$ and $\frac {1} {1/r_1 + 1/r_2} > R^*$} \STATE $R^*
\leftarrow \frac {1} {1/r_1 + 1/r_2}$, $L^* \leftarrow L$, $r_1^*
\leftarrow r_1$, $r_2^* \leftarrow r_2$\ \ENDIF \ENDFOR

 \end{algorithmic}
\end{minipage}
\end{algorithm}

\end{document}